# Methods for comparing rankings of search engine results

Judit Bar-Ilan[1]
Bar-Ilan University and The Hebrew University of Jerusalem, Israel
barilaj@mail.biu.ac.il

and

Mazlita Mat-Hassan
School of Computer Science and Information Systems
Birkbeck , University of London
azy@dcs.bbk.ac.uk

and

Mark Levene
School of Computer Science and Information Systems
Birkbeck, University of London
mark@dcs.bbk.ac.uk

**Abstract:** In this paper we present a number of measures that compare rankings of search engine results. We apply these measures to five queries that were monitored daily for two periods of about 21 days each. Rankings of the different search engines (Google, Yahoo and Teoma for text searches and Google, Yahoo and Picsearch for image searches) are compared on a daily basis, in addition to longitudinal comparisons of the same engine for the same query over time. The results and rankings of the two periods are compared as well.

---

[1] Corresponding author:
Judit Bar-Ilan
Department of Information Science
Bar-Ilan University
Ramat Gan, 52900, Israel
Tel: 972-3-5318351
Fax: 972-3-5353937
Email: barilaj@mail.biu.ac.il

**Introduction:**

In merely fifteen years the Web has grown to be one of the major information sources. Searching is a major activity on the Web [1,2], and the major search engines are the most frequently used tools for accessing information [3]. Because of the vast amounts of information, the number of results for most queries is usually in the thousands, sometimes even in the millions. On the other hand, user studies have shown [4-7] that users browse through the first few results only. Thus results ranking is crucial to the success of a search engine.

In classical IR systems, results ranking was based mainly on term frequency and inverse document frequency (see for example [8, pp. 29-30]). Web search results ranking algorithms take into account additional parameters such as the number of links pointing to the given page [9,10], the anchor text of the links pointing to the page, the placement of the search terms in the document (terms occurring in title or header may get a higher weight), the distance between the search terms, popularity of the page (in terms of the number of times it is visited), the text appearing in meta-tags [11], subject-specific authority of the page [12,13], recency in search index, and exactness of match [14].

Search engines compete with each other for users, and Web page authors compete for higher rankings with the engines. This is the main reason that search engine companies keep their ranking algorithms secret, as Google states [10]: "Due to the nature of our business and our interest in protecting the integrity of our search results, this is the only information we make available to the public about our ranking system …". In addition, search engines continuously fine-tune their algorithms in order to improve the ranking of the results. Moreover, there is a flourishing search engine optimization industry, founded solely in order to design and redesign Web pages so that they obtain high rankings for specific search terms within specific search engines (see for example Search Engine Optimization, Inc., www.seoinc.com/).

It is therefore clear from the above discussion that the top ten results retrieved for a given query have the best chance of being visited by Web users. This was the main motivation for the research we present herein, in addition to examining the changes over time in the top ten results for a set of queries of the largest search engines, which at the time of the first data collection were Google, Yahoo and Teoma (MSN search came out of beta on February 1, 2005 in the midst of the second round of data collection [15]). We also examined results of image searches on Google image search, Yahoo image search, and on Picsearch (www.picsearch.com/). The searches were carried out daily for about three weeks in October-November, 2004 and again in January-February, 2005. Five queries (3 text queries and 2 image queries) were monitored. Our aim was to study changes in the rankings over time in the results of the individual engines, and in parallel to study the similarity (or rather non-similarity) between the top ten results of these tools. In addition, we examined the changes in the results between the two search periods.

Ranking of search results is based on the problematic notion of relevance (for extended discussions see [16,17]. We have no clear notion of what is a "relevant document" for a given query, and the notion becomes even fuzzier when looking for "relevant documents" relating to the user's information seeking objectives. There are several transformations between the user's "visceral need" (a fuzzy view of the information problem in the user's mind) and the "compromised need" (the way the query is phrased taking into account the limitations of the search tool at hand) [18].

Some researchers (see for example [19]) claim that only the user with the information problem can judge the relevance of the results, while others claim that this approach is impractical (the user cannot judge the relevance of large numbers of documents) and suggest the use of judges or a panel of judges (e.g. in the TREC Conferences, the instructions for the judges appear in [20]). On the Web the question of relevance becomes even more complicated; users usually submit very short queries [4-7]. Consider, for example, the query "organic food". What kind of information is the user looking for: an explanation about what organic food is, a list of shops where organic food can be purchased (in which geographic location is the shop?), a site from which he/she can order organic food items, stories about organic food, medical evidence about the advantages of organic food, organic food recipes, and so on. What should the search engine return for such a query and how should it rank the results?

Most previous studies examining ranking of search results base their findings on human judgment. In an early study in 1998 by Su et al. [21], users were asked to choose and rank the five most relevant items from the first twenty results retrieved for their queries. In their study, Lycos performed better on this criteria than the other three search engines examined at the time. In 1999, Hawking et al. [22] evaluated the effectiveness of twenty public Web search engines on 54 queries. One of the measures used was the reciprocal rank of the first relevant document – a measure closely related to ranking. The results showed significant differences between the search engines and high inter-correlation between the measures. In 2002 Chowdhury and Soboroff [23] also evaluated search effectiveness based on the reciprocal rank – they computed the reciprocal rank of a known item for a query (a URL they apriori paired with the query). In a recent study in 2004, Vaughan [24] compared human rankings of 24 participants with those of three large commercial search engines, Google, AltaVista and Teoma, on four search topics. The highest average correlation between the human-based rankings and the rankings of the search engines was for Google, where the average correlation was 0.72. The average correlation for AltaVista was 0.49. Beg [25] compared the rankings of seven seven search engines on fifteen queries with a weighted measure of the users' behavior based on the order the documents were visited, the time spent viewing them and whether they printed out the document or not. For this study the results of Yahoo, followed by Google had the best correlation with this measure based on the user's behavior.

Other studies of search results rankings did not involve users. Soboroff et al. [26] based their study on the finding that differences in human judgments of relevance do not affect the relative evaluated performance of the different systems [27]. They proposed a ranking system based on randomly selecting "pseudo-relevant" documents. Zhao [28] submitted the query "cataloging department" to Google once a week for a period of ten weeks and studied the changes in the ranks of the twenty-four sites that were among the top-twenty pages during the data collection period. All but three web sites changed their position at least once during the observation period. The goal of Zhao's study was to try to understand how different parameters (e.g. PageRank, placement of keywords, structure of website) influence placement, and she provided descriptive statistics to that effect.

Fagin et al. [29] introduced a measure (described in the following section) to compare rankings of the top-k results of two search engines, even if the two lists of retrieved documents are not identical. The two lists may contain non-identical documents for two reasons: 1) since only the top-k results are considered, the search engine may have ranked the document after the k-th position, and 2) since the search engine has not indexed the given document (it is well-known that the overlap between the indexes of the different search engines was small at least in 1998-9, see [30,31]).

In a previous study [32], we compared the rankings of Google and AlltheWeb on several queries, by computing the size of the overlap, the Spearman correlation on the overlapping elements and a normalized Fagin measure. Each of these measures have their shortcomings (see next section), and thus besides the previous measures, we introduce herein an additional measure for comparing rankings. Two of the queries examined in this paper were also monitored in the previous work.

The goal of the current study is to examine changes in rankings of the top-ten results over time in a given search engine and to compare the rankings provided by different search engines using several comparison measures.

**The Measures**

We used four measures in order to assess the changes over time in the rankings of the search engines and to compare the results of the different search engines. The first and simplest measure is simply the *size of the overlap* between two top ten lists.

The second measure was Spearman's footrule [33,34]. Spearman's footrule is applied to two rankings of the same set; if the size of the set is $N$, all the rankings must be between 1 and $N$ (the measure is based on permutations, and thus no ties are allowed). Since the top ten results retrieved by two search engines for a given query, or retrieved by the same engine on two consecutive days are not necessarily identical, the two lists had to be transformed before Spearman's footrule could be computed. First the non-overlapping URLs were eliminated from both lists, and then the remaining lists were re-ranked; each URL was given its relative rank in the set of remaining URLs in each list. The result of the re-rankings are two permutations $\sigma_1$ and $\sigma_2$ on $1…S$, where $|S|$ is the number of overlapping URLs. After these transformations Spearman's footrule is computed as

$$Fr^{|S|}(\sigma_1, \sigma_2) = \sum_{i=1}^{|S|} |(\sigma_1(i) - \sigma_2(i))|.$$

When the two lists are identical, $Fr^{|S|}$ is zero, and its maximum value is ½$|S|^2$ when $|S|$ is even, and ½($|S|$+1)($|S|$-1) when $|S|$ is odd. If we divide the result by its maximum value, $Fr^{|S|}$ will be between 0 and 1, independent of the size of the overlap; we note that this is defined only for $|S|>1$. Thus we compute the *normalized Spearman's footrule, NFr*, for $|S|>1$

$$NFr = \frac{Fr^{(|S|)}}{\max Fr^{(|S|)}}$$

NFr ranges between 0 and 1; it attains the value 0 when the two lists are identically ranked and the value 1 when the lists appear in opposite order.

Our other measures are also in this range, but get the value 1 when the lists are identical and the value 0 when they are completely dissimilar. In order to be able to compare the results to those using other measures, we introduce $F$ as

$$F = 1 - NFr.$$

Note that Spearman's footrule is based on the re-ranked lists, and thus, for example, if the original ranks of the URLs that appear in both lists (i.e. the overlapping pairs) are (1,8), (2,9) and (3,10), the re-ranked pairs will be (1,1), (2,2) and (3,3) and the value of *Fr\** will be 1.

The third measure we utilized was one of the metrics introduced by Fagin et al. [29]. It is relatively easy to compare two rankings of the same list of items – for this, well-known statistical measures such as Kendall's tau, Spearman's rho or Spearman's footrule can easily be used. The problem arises when the two search engines that are being compared rank non-identical sets of documents. To cover this case (which is the usual case when comparing top-*k* lists created by different search engines), Fagin et al. [29] extended the previously mentioned metrics. Here we discuss only the extension of Spearman's footrule, but the extension of Kendall's tau is shown in their paper to be equivalent to the extension of Spearman's footrule. We note that a major point in their method was to develop measures that are either metrics or "near" metrics.

Spearman's footrule, is the $L_1$ distance between two permutations, given by $Fr(\sigma_1, \sigma_2) = \sum |\sigma_1(i) - \sigma_2(i)|$. This metric is extended to the case where the two lists are not identical, by assigning an arbitrary placement (which is larger than the length of the list) to documents appearing in one of the lists but not in the other; when comparing lists of length *k* this placement can be *k+1* for all the documents not appearing in the list. The rationale for this extension is that the ranking of those documents must be *k*+1 or higher, although Fagin et al. do not take into account the possibility that those documents are not indexed at all by the other search engine.

The extended metric now becomes
$$F^{(k+1)}(\tau_1, \tau_2) = 2(k-z)(k+1) + \sum_{i \in Z} |\tau_1(i) - \tau_2(i)| - \sum_{i \in S} \tau_1(i) - \sum_{i \in T} \tau_2(i)$$
where *Z* is the set of overlapping documents, *z* is the size of *Z*, *S* is the set of documents that are only in the first list, and *T* is the set of documents that appear in the second list only.

A problem with the measures proposed by Fagin et al. [29] is that when the two lists have little in common, the documents that are not common to the lists have a major effect on the measure. Our experiments show that usually the overlap between the top ten results of two search engines for an identical query is very small, and thus the non-overlapping elements have a major effect on the measure.

$F^{(k+1)}$ was normalized by Fagin et al. [29] so that the values lie between 0 and 1. For *k*=10 the normalization factor is 110. Thus we compute
$$G^{(k+1)} = 1 - \frac{F^{(k+1)}}{\max F^{(k+1)}}$$

which we refer to as the *G measure*.

As mentioned above, *Spearman's footrule* is calculated on the re-ranked list of overlapping elements, and ignores the actual rank of the overlapping elements. Thus for the case where there are only two overlapping elements, it cannot differentiate between the cases where the original placements of the overlapping elements are, say
1. (1,1), (2,2),
2. (1,9), (2,10), or
3. (1,2), (2,10).

In all three cases *F* is 1, since after the re-ranking in all three case we are considering the pairs (1,1) and (2,2).

The *G* measure *does* take into account the placement of the overlapping elements in the lists. For the above examples, the values of *G* will be:

1. 0.345
2. 0.055
3. 0.182

The *G* measure seems to capture our intuition that even though the overlapping elements appear in the same order in the two lists, if these appear in places which are more similar, the distance between the measures should be smaller. On the other hand, even if the top five documents are identical, and there is no additional overlap between the lists, the *G* measure will be 0.727 if the identical elements are in the same order and 0.618 if they appear in opposite order. I.e., the amount of change in *G* for a given overlap is rather small and is mainly determined by the size of the overlap.

For this reason we decided to experiment with an additional measure, which we call *M*. This measure tries to capture the intuition that identical or near identical rankings among the top documents (say the top three documents) is more valuable to the user than among the lower placed documents. First, let

$$M' = \sum_Z \left| \frac{1}{rank_1(i)} - \frac{1}{rank_2(i)} \right| + \sum_S \left( \frac{1}{rank_1(j)} - \frac{1}{11} \right) + \sum_T \left( \frac{1}{rank_2(j)} - \frac{1}{11} \right)$$

where *Z* is the set of the overlapping elements, $rank_1(i)$ is the rank of document i in the first set and $rank_1(i)$ is its rank in the second set (both ranks are defined for elements belonging to *Z*). In addition, *S* is the set of documents that appear in the first list but not in the second, while *T* is the set of elements that appear in the second list, but not in the first. These documents may appear in the other list as well, but their rank will be 11 or higher (since we consider the top-ten results only).; this is the reason that we subtract 1/11 from the reciprocal value of their rank. This measure differs from *G*, in that it gives a higher weight to higher ranking documents. We have to normalize this measure as well, and to make sure that for identical lists the value of the measure is 1 and for lists, where |Z|=10 and the documents appear in opposite order, the value is 0. Thus we let

$$M = 1 - \frac{M'}{4.03975}.$$

To demonstrate the difference between the emphases of *G* and *M*, assume that the two lists are identical, except that

a) the first document is different in the two lists. In this case *G* will be 0.818 and *M* will be 0.5499.

b) the last document is different in the two lists. In this case *G* will be 0.9818 and *M* will be 0.9955.

Note that in both cases the *F*-value will be 1. Let us compute the values of *M* for the examples we used for comparing *F* with *G*, i.e.

1. (1,1), (2,2),
2. (1,9), (2,10), and
3. (1,2), (2,10)

Table 1: Comparing *F*, *G* and *M*

|            | F | G     | M     |
|------------|---|-------|-------|
| (1,1), (2,2) | 1 | 0.345 | 0.653 |
| (1,9), (2,10) | 1 | 0.055 | 0.015 |
| (1,2), (2,10) | 1 | 0.182 | 0.207 |

From Table 1, we can see that in the first case, when the overlapping elements are in high positions in both sets, *M* is considerably higher than *G*. On the other hand, when the overlapping elements are in top ranks in the first list but appear at the bottom of the second list, *M* is much lower than *G*. Thus we see that *M* captures our intuition and gives higher weights to higher ranking overlapping elements.

**Data collection**

The data collection for the first round was carried out by students. Their assignment involved picking a text query and a picture query from a given list of queries and to submit these queries to the appropriate search engine once a day for a period of fourteen days. The students started data collection at different dates; therefore if two or more students monitored the same query, we had data for these queries for more than fourteen days. Sometimes the students skipped a day of data collection, or there was no overlap between the students work, and thus the data for the first period is not completely continuous. During the second period all the queries were monitored for 21 consecutive days. Table 2 displays the queries, the number of days these queries were monitored and the time span of the data collection for each period.

The queries submitted to the search engines were carefully chosen. For the text queries, we chose a query to represent a topical topic (US elections 2004) and two queries from our previous study [32]; organic food and DNA evidence. These two queries were chosen as we were interested to investigate whether previously monitored URLs still available during the current observation period. For image queries, the queries were chosen to represent *places* (Bondi beach and Twin towers) and an event (Twin towers). The query "Twin towers" was particularly interesting as it represents both, a place and an event (the 9/11 attacks on the World Trade Centre).

Table 2: Data collection

| Query | # days, first period | dates first period | # days, second period | dates second period |
|---|---|---|---|---|
| 1. US elections 2004 | 9 | 1-15 November, 2004 | 21 | 24 January – 13 February, 2005 |
| 2. DNA evidence | 21 | 22 October – 11 November, 2004 | 21 | 24 January – 13 February |
| 3. organic food | 17 | 23 October – 8 November, 2004 | 21 | 24 January – 13 February |
| 4. Twin towers | 24 | 22 October – 15 November, 2004 | 21 | 24 January – 13 February |
| 5. Bondi beach | 18 | 22 October – 8 November, 2004 | 21 | 24 January – 13 February |

The first three queries were text searches and were submitted at each data collection point to Google, Yahoo and Teoma, while the last two queries were image searches and were submitted to Google image search, Yahoo image search and to Picsearch (www.picsearch.com). The URLs and the rankings of the top ten results for each query and for each search engine were recorded at each data collection point. For the image searches, the URLs of the images (and not of the embedding pages) were recorded.

**Data analysis**

For a given search engine and a given query we computed the overlap (*O*), Spearman's footrule (*F*), Fagin's measure (*G*) and our new measure (*M*), on the results for consecutive data collection points. The results of pairs of engines were also compared by computing the same measures for the two ranked lists retrieved by the two search engines on the same day, for each day recorded. The two periods were compared on all five queries; we calculated the overlap between the two periods and assessed the changes in the rankings of the overlapping elements based on the average rankings. For all the queries, the maximum values of all the measures were 1, and 10 for overlap, except for "US elections 2004", where the maximum overlap was only 9 for Google, noting that, at times, the data was not collected on consecutive days. An additional reason for this could be that the data was created very close to the elections (between November 1 and 15, 2004; the elections were held on November 2, 2004).

**Results**
*The first round*

**Table 3: Measures for the changes in ranking of the individual engines over time – round 1**

| query + search engine | overlap | | F | | G | | M | | #URLs | overlap between first and last day |
|---|---|---|---|---|---|---|---|---|---|---|
| | avg | min | avg | min | avg | min | avg | min | | |
| **US elections 2004** | | | | | | | | | | |
| google | 8.13 | 7 | 0.86 | 0.51 | 0.78 | 0.6 | 0.74 | 0.74 | 18 | 8 |
| yahoo | 9.25 | 8 | 0.95 | 0.84 | 0.92 | 0.8 | 0.95 | 0.86 | 15 | 8 |
| teoma | 10 | 10 | 1 | 1 | 1 | 1 | 1 | 1 | 10 | 10 |
| **organic food** | | | | | | | | | | |
| google | 10 | 10 | 0.95 | 0.80 | 0.98 | 0.91 | 0.99 | 0.95 | 10 | 10 |
| yahoo | 10 | 10 | 1 | 1 | 1 | 1 | 1 | 1 | 10 | 10 |
| teoma | 9.94 | 9 | 1 | 1 | 0.99 | 0.95 | 0.99 | 0.98 | 11 | 9 |
| **DNA evidence** | | | | | | | | | | |
| google | 9.1 | 8 | 0.98 | 0.88 | 0.93 | 0.84 | 0.97 | 0.91 | 18 | 7 |
| yahoo | 10 | 10 | 1 | 1 | 1 | 1 | 1 | 1 | 10 | 10 |
| teoma | 9.63 | 9 | 0.95 | 0.80 | 0.94 | 0.85 | 0.96 | 0.84 | 12 | 9 |
| **Twin towers** | | | | | | | | | | |
| google | 9.28 | 7 | 0.88 | 0.5 | 0.89 | 0.62 | 0.92 | 0.70 | 13 | 7 |
| yahoo | 9.63 | 8 | 0.94 | 0.56 | 0.95 | 0.78 | 0.95 | 0.75 | 14 | 10 |
| picsearch | 9.78 | 5 | 1 | 1 | 0.98 | 0.67 | 0.99 | 0.85 | 14 | 5 |
| **Bondi beach** | | | | | | | | | | |
| google | 8.06 | 7 | 0.81 | 0.43 | 0.84 | 0.62 | 0.89 | 0.74 | 13 | 10 |
| yahoo | 8.76 | 2 | 0.92 | 0 | 0.86 | 0.15 | 0.84 | 0.05 | 21 | 2 |
| picsearch | 9.94 | 9 | 0.99 | 0.96 | 0.98 | 0.82 | 0.98 | 0.78 | 11 | 9 |

Google's set of results and rankings fluctuated slightly during the period of data collection, with the exception of the query "organic food" that was very stable (the same is true of the other two search engines for this query). Interestingly, even though Google covered 13 URLs among the top-ten results for the query "Bondi beach", the result sets for the first and last day were identical. Figure 1 depicts the changes in the placements and occurrence of the URLs during the data collection period for the query "DNA evidence". We see from the figure, that the top-three places were stable during the whole period. URL4 (http://books.nap.edu/html/DNA/) was ranked fourth for eight days, then appeared in the top-ten for five additional days (in places five and six) and then disappeared from the top-ten (although it continued to exist, and as we shall see later, it reappeared in the top-ten during the second data collection period). The fourth place was taken by URL5 (www.nap.edu/catalog/5141.html), that was initially ranked number 5. Both pages contain information on a 1996 publication of the Committee of DNA Forensic Science of the US National Research Council, although URL4 has much more actual content, while URL5 offers the purchase of the report. URL1 (www.howstuffworks.com/dna-evidence.htm) presents popular information, and is identical in content to URL15, which appeared as number 4 on days 19 and 20. URLs 2 and 3 (www.ojp.usdoj.gov/nij/dna and www.ojp.usdoj.gov/nij/dna_evbro/) contain information provided by the US Department of Justice on the topic.

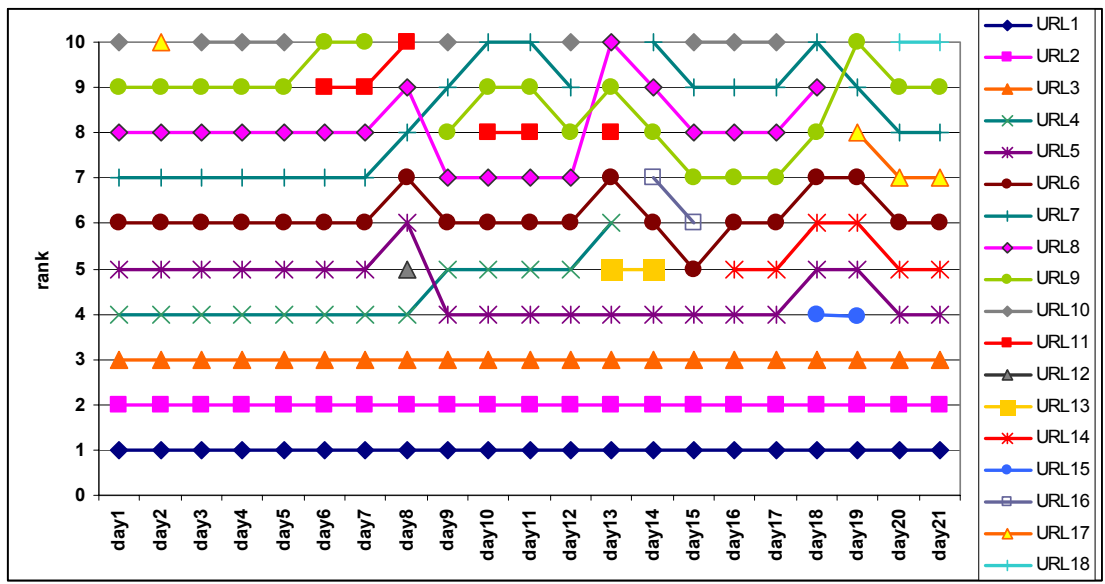

**Figure 1: The top-ten results of Google for the query "DNA evidence"**

Yahoo showed either minor or no changes in the results of the text queries, and was also relatively stable on the image query "Twin towers", however its results fluctuated heavily for the query "Bondi beach" – only two URLs appeared among the top-ten for all 18 days. We could identify four different sub-periods, as can be seen in Table 4.

Teoma was highly stable during the period. Picsearch retrieved almost identical results on "Bondi beach" at each data collection point, while for the query "Twin towers" there was a single, but considerable change on November 10, 2004, otherwise the results and rankings were stable.

**Table 4: Yahoo's top-ten results for the image query "Bondi beach"**

|       | days 1-3 | day 4 | days 5-11 | days 12-18 |
|-------|----------|-------|-----------|------------|
| URL1  | 1        |       | 1         |            |
| URL2  | 2        |       | 2         |            |
| URL3  | 3        | 5     |           |            |
| URL4  | 4        | 7     |           |            |
| URL5  | 5        | 6     | 5         | 6          |
| URL6  | 6        | 8     | 3         | 8          |
| URL7  | 7        |       | 4         |            |
| URL8  | 8        |       | 10 (was #9 on day 7) |  |
| URL9  | 9        | 10    | 7         |            |
| URL10 | 10       |       | 6         |            |
| URL11 |          | 1     |           | 3          |
| URL12 |          | 2     |           | 1          |
| URL13 |          | 3     |           | 2          |
| URL14 |          | 4     |           | 4          |
| URL15 |          | 9     | 8         |            |
| URL16 |          |       | 9 (wasn't among the top-ten on day 7) | |
| URL17 |          |       | #10 on day 7 |         |
| URL18 |          |       |           | 5          |
| URL19 |          |       |           | 7          |
| URL20 |          |       |           | 9          |
| URL21 |          |       |           | 10         |

We also compared the rankings of the different search engines on the same query on the same day, using the same measures. There was no overlap between the top-ten results of any of the pairs of the search engines for the image query "Bondi beach". The situation was almost the same for the image query "Twin towers", except for a single image that appeared as #1 one Google during the whole period, and fluctuated between places 7 to 9 on Yahoo. Therefore, Table 5 displays the measures (average, minimum and maximum values) for all the text queries and for the image query "Twin towers" for the pair Google-Yahoo only.

**Table 5: Measures for comparing the rankings of the different search engines on identical queries at the same data collection points – first round**

| query + search engine | overlap | | | *F* | | | *G* | | | *M* | | |
|---|---|---|---|---|---|---|---|---|---|---|---|---|
|  | avg | min | max | avg | min | max | avg | min | max | avg | min | max |
| **US elections 2004** | | | | | | | | | | | | |
| google-yahoo | 3.44 | 3 | 4 | 0.43 | 0.11 | 0.56 | 0.36 | 0.29 | 0.44 | 0.29 | 0.25 | 0.35 |
| yahoo-teoma | 2.78 | 2 | 3 | 1 | 1 | 1 | 0.28 | 0.2 | 0.33 | 0.25 | 0.21 | 0.27 |
| google-teoma | 4.78 | 4 | 6 | 0.63 | 0.44 | 0.75 | 0.42 | 0.36 | 0.47 | 0.36 | 0.34 | 0.37 |
| **organic food** | | | | | | | | | | | | |
| google-yahoo | 7 | 7 | 7 | 0.52 | 0.5 | 0.58 | 0.61 | 0.56 | 0.69 | 0.51 | 0.48 | 0.55 |
| yahoo-teoma | 2 | 2 | 2 | 1 | 1 | 1 | 0.31 | 0.31 | 0.31 | 0.32 | 0.32 | 0.32 |
| google-teoma | 3 | 3 | 3 | 0.88 | 0.5 | 1 | 0.28 | 0.24 | 0.31 | 0.26 | 0.22 | 0.29 |
| **DNA evidence** | | | | | | | | | | | | |
| google-yahoo | 3.95 | 3 | 4 | 1 | 1 | 1 | 0.50 | 0.45 | 0.53 | 0.66 | 0.65 | 0.67 |
| yahoo-teoma | 3 | 3 | 3 | 0.5 | 0.5 | 0.5 | 0.33 | 0.31 | 0.35 | 0.51 | 0.50 | 0.52 |
| google-teoma | 1 | 1 | 1 | N/A | N/A | N/A | 0.18 | 0.18 | 0.18 | 0.45 | 0.45 | 0.45 |
| **Twin towers** | | | | | | | | | | | | |
| google-yahoo | 1 | 1 | 1 | N/A | N/A | N/A | 0.05 | 0.04 | 0.09 | 0.02 | 0.01 | 0.04 |

The highest overlap (7 overlapping URLs in the top-ten) was between Google and Yahoo for the query "organic food". The similarity measures are not very high, because the relative ranking of these URLs by the two engines are somewhat different: #1 on Google is #2 on Yahoo and visa versa, #4 on Google is #8 on Yahoo (and #4 on Yahoo is between the 7$^{th}$ and 10$^{th}$ places on Google). Even though there are only two overlapping elements for the query "organic food" between Yahoo and Teoma, their rankings are the same (#2 and #3 on both engines), thus the *F* value is one. The *G* and *M* values are relatively low because the size of the overlap is small, *M* is slightly higher than *G*, because the ranks of the overlapping elements are relatively high. These two URLs also appear on Google's lists: #2 on Teoma and Yahoo is #1 on Google, and #3 on Teoma and Yahoo ranks between 5 and 8 on Google.

There are two cases where the size of the overlap is 1 (Yahoo-Teoma for "DNA evidence", and Google-Yahoo for "Twin towers"): the considerable difference between the *G* values for these two cases are caused because of the different ranks of the overlapping element. For "DNA evidence" both engines ranked the overlapping URL as #1, while for "Twin towers", Google ranked the overlapping URL as #1, and Teoma's rank for this URL varies between 6 and 9.

Let us take a closer look at the query "DNA evidence". All three engines agree on the top-ranked URL (www.howstuffworks.com/dna-evidence.htm) for the whole period. Google and Yahoo overlap on four URLs, except for a single day where there were only three overlapping URLs. These URLs are constant during the whole period: the first three are URLs 1-3 of Google (mentioned above) – they are ranked 1, 3 and 4 on Yahoo respectively. We see that there is a high degree of agreement about the top results between the two search engines. The fourth overlapping URL is #7 on Yahoo and fluctuates between ranks 7 and 10 on Yahoo. Rank 2 on Yahoo (www.ncjrs.org/txtfiles/dnaevid.txt) is ranked as #10 on Teoma (and is not among the top-ten in Google). Teoma overlaps with Yahoo on URLs ranked 1, 2 and 5 on Yahoo's lists. These URLs are ranked 1, 10 and 4-5 on Teoma.

*The second round*

Table 6 summarizes the findings related to the rankings of each search engine over time. For each measure we provided the average and the minimum (over 21 days). The maximum value attained for each *F*, *G* and *M* was 1, and for the overlap 10.

Unlike in the first round, this time the results for the query "US elections 2004" were rather stable for all the engines. This finding is not surprising, since the second round took place almost three months after the elections.

**Table 6: Measures for the changes in ranking of the individual engines over time – round 2**

| query + search engine | overlap | | F | | G | | M | | #URLs located (whole period) | overlap between first and last day |
|---|---|---|---|---|---|---|---|---|---|---|
| | avg | min | avg | min | avg | min | avg | min | | |
| **US elections 2004** | | | | | | | | | | |
| google | 9.65 | 8 | 0.9 | 0.3 | 0.94 | 0.65 | 0.88 | 0.31 | 12 | 9 |
| yahoo | 9.85 | 8 | 1 | 1 | 0.98 | 0.82 | 0.99 | 0.79 | 13 | 8 |
| teoma | 9.75 | 8 | 0.98 | 0.85 | 0.97 | 0.8 | 0.97 | 0.8 | 13 | 9 |
| **organic food** | | | | | | | | | | |
| google | 9.1 | 5 | 0.93 | 0.67 | 0.92 | 0.46 | 0.95 | 0.57 | 15 | 8 |
| yahoo | 9.9 | 9 | 1 | 1 | 0.99 | 0.91 | 0.99 | 0.96 | 11 | 10 |
| teoma | 10 | 10 | 1 | 1 | 1 | 1 | 1 | 1 | 10 | 10 |
| **DNA evidence** | | | | | | | | | | |
| google | 9.05 | 8 | 0.99 | 0.9 | 0.94 | 0.84 | 0.98 | 0.9 | 19 | 7 |
| yahoo | 9.97 | 8 | 1 | 1 | 0.99 | 0.85 | 0.99 | 0.92 | 12 | 9 |
| teoma | 9.65 | 8 | 0.98 | 0.88 | 0.97 | 0.82 | 0.98 | 0.87 | 14 | 8 |
| **Twin towers** | | | | | | | | | | |
| google | 9.65 | 6 | 0.96 | 0.44 | 0.96 | 0.6 | 0.96 | 0.45 | 15 | 6 |
| yahoo | 8.75 | 5 | 0.9 | 0.3 | 0.83 | 0.42 | 0.74 | 0.15 | 20 | 7 |
| picsearch | 9.8 | 6 | 1 | 1 | 0.98 | 0.55 | 0.99 | 0.72 | 14 | 6 |
| **Bondi beach** | | | | | | | | | | |
| google | 9.55 | 7 | 0.99 | 0.75 | 0.95 | 0.69 | 0.97 | 0.82 | 15 | 9 |
| yahoo | 8.65 | 5 | 0.91 | 0 | 0.85 | 0.31 | 0.81 | 0.14 | 20 | 6 |
| picsearch | 9.95 | 9 | 0.99 | 0.96 | 0.99 | 0.91 | 0.99 | 0.96 | 11 | 9 |

Teoma retrieved exactly the same results in the same order on all 21 days for the query "organic food". On the other hand, Yahoo image searches had the most fluctuations. Figure 3 depicts the fluctuations in ten out of the twenty URLs that were identified by Yahoo for the query "Bondi beach".

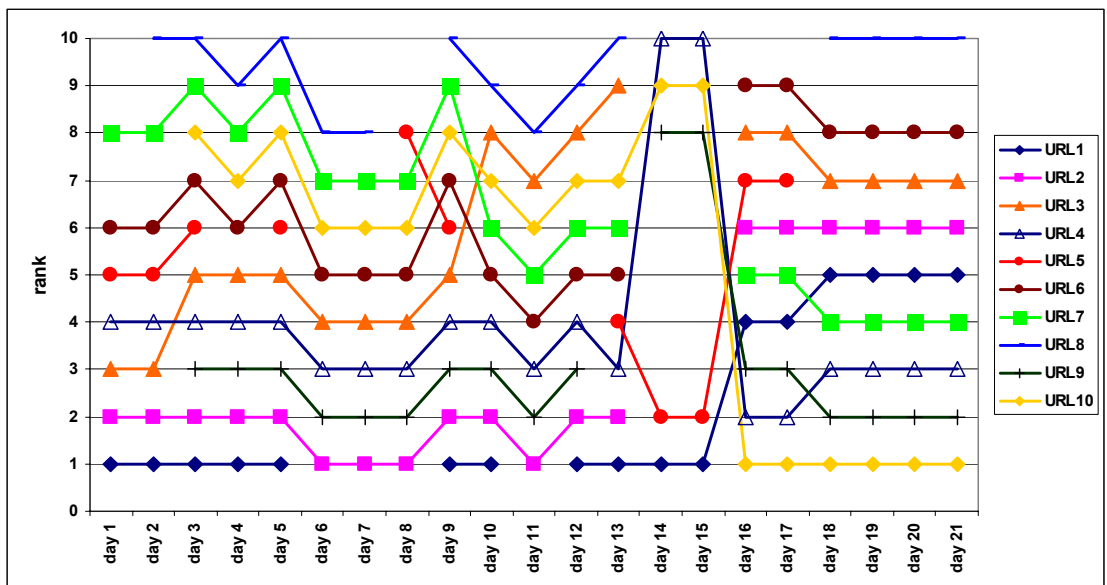

**Figure 2: Fluctuations in rankings of Yahoo for the query "Bondi beach"**

Next we compared the rankings of the different search engines on the same query on the same day. This time, there was no overlap at all between the search engine results for the image queries. Therefore, Table 7 displays the measures (average, minimum and maximum values) for the text queries only.

**Table 7: Measures for comparing the rankings of the different search engines on identical queries at the same data collection point**

| query + search engine | overlap | | | F | | | G | | | M | | |
|---|---|---|---|---|---|---|---|---|---|---|---|---|
| | avg | min | max | avg | min | max | avg | min | max | avg | min | max |
| **US elections 2004** | | | | | | | | | | | | |
| google-yahoo | 4 | 3 | 5 | 0.19 | 0 | 0.5 | 0.46 | 0.29 | 0.55 | 0.33 | 0.14 | 0.47 |
| yahoo-teoma | 5.0 | 4 | 6 | 0.34 | 0.11 | 0.75 | 0.44 | 0.31 | 0.49 | 0.31 | 0.14 | 0.38 |
| google-teoma | 5.2 | 4 | 6 | 0.58 | 0.5 | 0.78 | 0.46 | 0.4 | 0.58 | 0.35 | 0.27 | 0.45 |
| **organic food** | | | | | | | | | | | | |
| google-yahoo | 6.4 | 5 | 7 | 0.44 | 0.42 | 0.67 | 0.53 | 0.45 | 0.55 | 0.46 | 0.33 | 0.48 |
| yahoo-teoma | 2 | 2 | 2 | 0 | 0 | 0 | 0.26 | 0.25 | 0.27 | 0.18 | 0.16 | 0.2 |
| google-teoma | 3.1 | 2 | 4 | 0.36 | 0 | 0.75 | 0.22 | 0.18 | 0.31 | 0.11 | 0.09 | 0.14 |
| **DNA evidence** | | | | | | | | | | | | |
| google-yahoo | 4 | 4 | 4 | 0.75 | 0.75 | 0.75 | 0.45 | 0.42 | 0.46 | 0.61 | 0.6 | 0.62 |
| yahoo-teoma | 5 | 5 | 5 | 0.67 | 0.67 | 0.67 | 0.53 | 0.45 | 0.55 | 0.61 | 0.56 | 0.63 |
| google-teoma | 2 | 2 | 2 | 1 | 1 | 1 | 0.38 | 0.29 | 0.39 | 0.52 | 0.5 | 0.53 |

Let us examine two cases more closely, Google vs. Teoma on "organic food", where the *M* values are lower than the *G* values; and Yahoo vs. Teoma on "DNA evidence" where the *M* values are higher than the *G* values.

For the query "organic food" the number of overlapping URLs for Google and Teoma varies between 2 and 4. Two URLs overlap on all days, except one: www.organicfood.co.uk/, which is #1 on Google, and #4 on Teoma (and #2 on Yahoo) and www.rain.org/~sals/my.html, which is between #6 and #9 on Google (and not among the top ten on day 2) and #3 on Teoma (its rank varies between 3 and 4 on Yahoo as well). Thus the *F* values, based on relative rankings only, are low, the *M* value is very low as well, because the number of overlapping URLs is small, and there is considerable disagreement between the rankings, while the *G* value is higher, because it puts more weight on the number of overlapping elements, and less on their relative placements.

We identified five overlapping URLs for Yahoo and Teoma for the query "DNA evidence". The overlapping URLs are ranked 1, 2, 3, 7 and 8 by Yahoo and 1, 5-7, 4-5, 2 and 8-10 by Teoma, respectively. The *M* value is relatively high, because the top elements overlap, even if the relative rankings are not the same for the two engines.

### *Comparison between the rounds*

The results for the first round were collected at the end of October until the beginning of November, 2004. The second round took place three months later at the end of January until the beginning of February, 2005. In this section we examine how the top-ten results changed in three months. The aggregated results are displayed in Table 8.

**Table 8: Changes to the top-ten results between the rounds**

| | query | US elections 2004 | organic food | DNA evidence | Twin towers | Bondi beach |
|---|---|---|---|---|---|---|
| **Google** | # URLs identified during both periods | 19 | 15 | 26 | 20 | 17 |
| | overlap | 11 | 10 | 13 | 8 | 11 |
| | # URLs in first set, but missing from second set | 7 | 0 | 5 | 4 | 2 |
| | min change in average ranking | 0.24 | 0 | 0 | 0.34 | 0 |
| | max change in average ranking | 2.33 | 7.29 | 4 | 4.04 | 2.77 |
| **Yahoo** | # URLs identified during both periods | 18 | 13 | 12 | 24 | 34 |
| | overlap | 9 | 8 | 10 | 8 | 8 |
| | # URLs in first set, but missing from second set | 5 | 2 | 0 | 4 | 13 |
| | min change in average ranking | 0 | 0 | 0 | 1.53 | 1.72 |
| | max change in average ranking | 3.63 | 2.1 | 5 | 5.16 | 4.73 |
| **Teoma** | # URLs identified during both periods | 15 | 12 | 21 | | |
| | overlap | 8 | 9 | 5 | | |
| | # URLs in first set, but missing from second set | 2 | 2 | 7 | | |
| | min change in average ranking | 0 | 0 | 0 | | |
| | max change in average ranking | 4.38 | 2 | 3.76 | | |
| **Picsearch** | # URLs identified during both periods | | | | 25 | 18 |
| | overlap | | | | 3 | 4 |
| | # URLs in first set, but missing from second set | | | | 11 | 7 |
| | min change in average ranking | | | | 3.14 | 0.63 |
| | max change in average ranking | | | | 6 | 5.78 |

The average rank of a URL for a search engine in a search round is the sum of the rankings it received on each day the URL appeared among the top-ten results of the search engine for the query, divided by the number of days it appeared in the top-ten list. Thus the average rank of a URL is between 1 and 10. The change in the average rank of a URL is defined as the absolute value of the difference between its average rank in round 1 and round 2 (the value is undefined if it was missing from either round). The minimum and maximum values were computed for each search engine and each query over all the URLs for which the change was defined. The smaller the maximum change, the more similar are the rankings in the two rounds.

Here we see again, that the results of Teoma and Yahoo (text searches) were most stable (the number of URLs identified in the first round, but missing in the second round from the top-ten, were the smallest). The query "organic food" had more stable results than the other queries we examined (least number of URLs identified, and least number of URLs missing from the second set).

At this point in time, image searches are rather different from text searches. Even though the results of Picsearch were very stable during the each data collection period, the results changed considerably between periods. Google was most stable for image searches. Google admitted in November 2004 that it had not updated its image database for some time [35]. In spite of this report, we still observed considerably changes during the first round in the top-ten results of Google for our queries. On February 8, 2004, Google announced that it has refreshed and expanded its image database [36]. This was in the middle of the second round of data collection. We saw some changes in the top-ten results for Google on February 9, 2005. The overlap with the results of the previous day was only six URLs for "Twin towers", and their relative rankings also changed considerably. On the other hand only a minor change was observed for "Bondi beach", and we saw more considerable changes in the daily results in the top-ten set before the date of the expansion.

For the text queries, in eight out of nine cases (three search engines, three queries each) there was at least one URL whose average rank has not changed between the search rounds (as can be seen from the rows for minimum change in average ranking). In all of these cases, this minimum was achieved by the top ranking URL, i.e., the top-ranking URL was ranked #1 at each data collection point both in round one and in round two. For the image searches, only for Google, for the query "Bondi beach", the #1 URL remained the same during both periods.

**Discussion**

The queries "DNA evidence" and "organic food" were also monitored in our previous study [32]. Then we submitted the queries to Google and to AllTheWeb, during two data collection periods: in October 2003 and in January 2004 (i.e., exactly a year before the current data collection rounds). We identified 4 URLs for the query "DNA evidence" and 6 URLs for "organic food" that appeared in all four data collection rounds. Figures 3 and 4 depict the average rankings of these URLs during the four data collection periods. The rankings for "DNA evidence" are much more stable than the rankings for "organic food". It is interesting to see that so many URLs remained in the top-ten results for these queries for over a year.

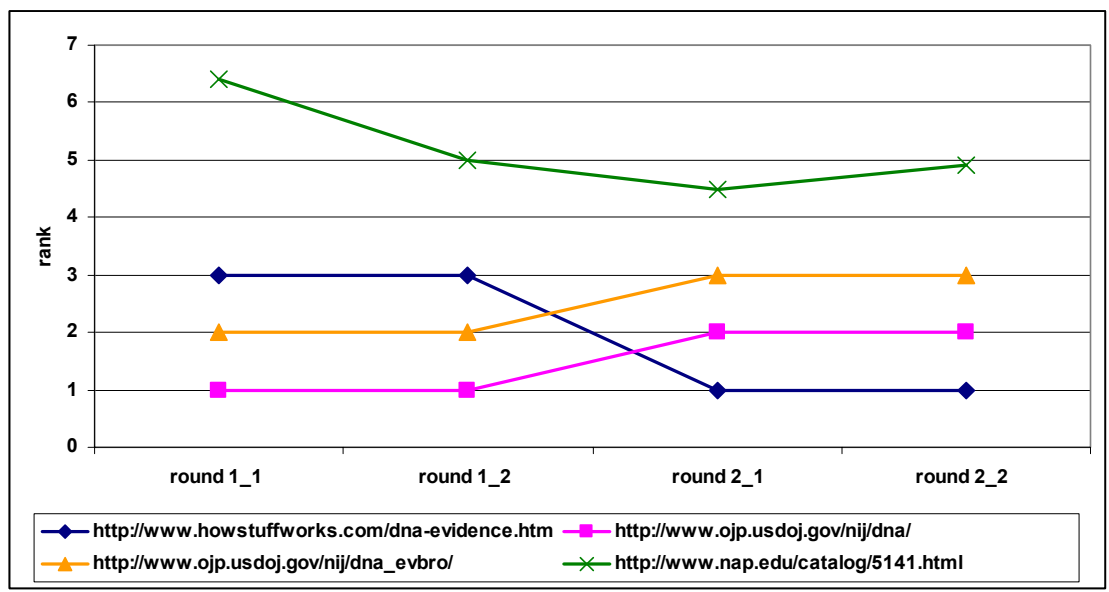

**Figure 3: The changes to the average rankings assigned by Google in the four data collection periods for the query "DNA evidence"**

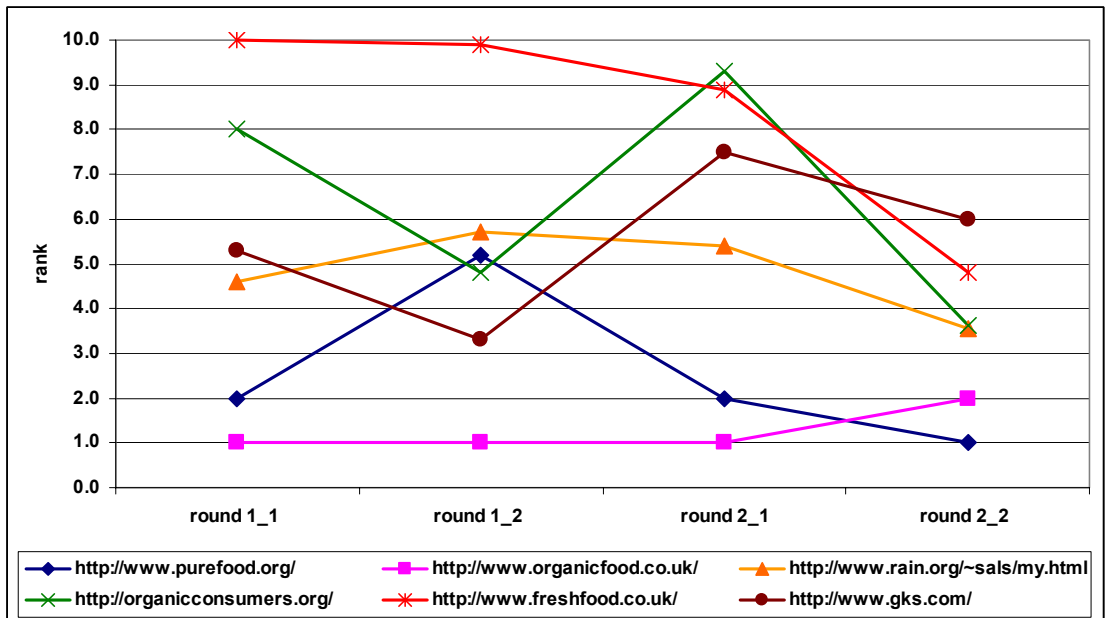

**Figure 4: The changes to the average rankings assigned by Google in the four data collection periods for the query "organic food"**

Jux2 (www.jux2.com/), a tool for visualizing the overlap between the top results of Google, Yahoo and AskJeeves, reports that on the 500 most popular search terms, the average overlap between Google and Yahoo was 3.8, and for 30% of the queries the overlap was between 0 and 2 [37]. It also reports that the overlap between Google and AskJeeves (powered by Teoma) is even smaller, 3.4 on average, while the average overlap between Yahoo and AskJeeves is only 3.1 on average. For "our" three test queries, the average overlap was 4.8 between Google and Yahoo, 3.4 between Google and Teoma, and 4 between Yahoo and Teoma. Thus our results slightly differ from the statistics provided by Jux2. They tested a much larger set of queries, and most probably the queries observed by us were not among the ones they checked. Jux2 do not provide the methods they used for measuring the similarity in rankings; they only provide the descriptive statistics.

In this study we also experimented with image queries, an extension to our previous study where only text queries were observed. The queries were specifically chosen to represent a *place* (Bondi beach, Twin towers) and an *event* (Twin towers). Our experiments have shown that while results were very stable during each specific data collection for all search engines involved, the results changed considerably between periods. Overlapping between search engines was non-existent or very minimal. On April 17, 2005 we queried the search engines Google and Yahoo for the existence of URLs identified by Yahoo and Google, respectively, during the second search round for the queries "DNA evidence" and "Bondi beach" (for the image query we submitted the URLs in which the images were embedded). Google indexed all 12 URLs identified by Yahoo for "DNA evidence", but only 7 out of the 20 URLs located by Yahoo with images on "Bondi beach". Yahoo covered Google's image searches much better; it indexed 11 out of the 15 URLs located by Google, but did worse on the text query, where it indexed 14 out of the 19 URLs identified during the period by Google on the query "DNA evidence". Note, that we checked the overlap after the latest announced update of Google Images [36], but this update took place in the middle of the second data collection round, thus the URLs located in the second round were partially collected from the new database.

**Conclusions**

In this paper, we experimented with a number of measures in order to assess the changes that occur over time to the rankings of the top ten results of search engines, and to assess the differences in the rankings of different search engines. In our previous study, we computed the overlap, Spearman's rho and Fagin's *G* measure. We observed that these measures are not fully satisfactory on their own, and thus we recommended that all of the three measures should be used.

In the current study we computed four measures: the overlap, Spearman's footrule, *F*, Fagin's *G* measure, and the new *M* measure. Our reason for introducing this new measure was to minimize the problems related to the other measures. The overlap ignores rankings altogether, Spearman's footrule is based only on the relative rankings and ignores the non-overlapping elements completely, and, finally, Fagin's measure gives far too much weight to the size of the overlap. The new measure attempts to take into account both the overlapping and the non-overlapping elements, and gives higher weight to the overlapping URLs among the top-ranking results. It seems that the *M* measure better captures our intuition about the quality of rankings, but further studies are needed to show the full utility of this measure (and/or experimenting with additional measures).

We experimented both with text and image queries. Results of image queries were less stable, and the overlap between the results of the different search tools was non-existent or minimal. This is striking compared to the average overlap of 4.1 between all pairs of search engines for all the text queries. Thus it seems that either there is much more agreement on the "importance" of textual data versus image data,or that the image databases of the different search engines are almost disjoint.

Our results seem to indicate that even though the overlap between the top-ranked documents for the image queries is lower than for text queries, the overlap is still considerable. Thus it seems that the differences in the coverage of the image databases only provide a partial explanation for the different results obtained by the different search tools for the image queries. Further studies are needed in this area as well.